\pdfoutput=1
\documentclass[]{trbunofficial}
\usepackage{graphicx}
\usepackage[hidelinks]{hyperref}
\usepackage{booktabs}
\usepackage{multirow}
\usepackage{caption}
\usepackage{subcaption}

\AuthorHeaders{Mehrotra, Hunter, Konishi, Akash, Zhang, Misu, Kumar, Reid, Jain}
\title{Trust in shared automated vehicles - Study  on two mobility platforms}

\author{%
  \textbf{Shashank Mehrotra}\\
  Honda Research Institute USA Inc.\\
  San Jose, CA, 95134\\
  Email: shashank\_mehrotra@honda-ri.com\\
  \hfill\break%
  \textbf{Jacob G Hunter }\\
  School of Mechanical Engineering\\
  Purdue University, West Lafayette, IN 47907-2088\\
  Email: hunte168@purdue.edu\\
  \hfill\break%
  \textbf{Matthew Konishi}\\
  School of Mechanical Engineering\\
  Purdue University, West Lafayette, IN 47907-2088\\
  Email: mkonish@purdue.edu\\
  \hfill\break%
  \textbf{Kumar Akash}\\
  Honda Research Institute USA Inc.\\
  San Jose, CA, 95134\\
  Email: kakash@honda-ri.com\\
  \hfill\break%
  \textbf{Zhaobo Zheng}\\
  Honda Research Institute USA Inc.\\
  San Jose, CA, 95134\\
  Email: zhaobo\_zheng@honda-ri.com\\
  \hfill\break%
  \textbf{Teruhisa Misu}\\
  Honda Research Institute USA Inc.\\
  San Jose, CA, 95134\\
  Email: tmisu@honda-ri.com\\
  \hfill\break%
  \textbf{Anil Kumar}\\
  San Jose State University\\
  San Jose, California 95192-0085\\
  Email: anil.kumar@sjsu.edu\\
  \hfill\break%
  \textbf{Tahira Reid}\\
  School of Mechanical Engineering\\
  Purdue University, West Lafayette, IN 47907-2088\\
  Email: tahira@purdue.edu\\
  \hfill\break%
  \textbf{Neera Jain}\\
  School of Mechanical Engineering\\
  Purdue University, West Lafayette, IN 47907-2088\\
  Email: neerajain@purdue.edu
  }

\begin{document}
\maketitle
\section{Abstract}

The ever-increasing adoption of shared transportation modalities across the platforms, has the potential to change the preferences and usage of different mobilities fundamentally. It also raises several challenges to designing and developing automated mobilities that enable a large population to take advantage of this emergent technology. One such challenge is the need to understand how trust in one automated mobility may impact trust in another. With this understanding, it is easier for researchers to determine whether future mobility solutions will have acceptance within different population groups. This study focuses on identifying the differences in trust across other mobility and how trust evolves across their use for participants who preferred an aggressive driving style. A dual mobility simulator study was designed in which 48 participants experienced two automated mobilities (car and sidewalk). The results found that participants showed increased levels of trust when they transitioned from car to sidewalk mobility.The findings from the study help inform and identify how people can develop trust in future mobility platforms,and could inform the design of interventions that may help improve the trust and acceptance of future mobility.

\hfill\break%
\noindent\textit{Keywords}: Trust in automation, shared automated vehicles, dual-mobility experiment
\newpage

\section{Introduction}

As defined by \citeauthor{shaheen_shared_2016} \cite{shaheen_shared_2016}, ``Shared mobility is the shared use of a vehicle, bicycle, or other low-speed mode that enables users to have short-term access to transportation modes on an `as needed' basis''. Typical services include car sharing, bike sharing, scooter sharing, on-demand ride services, ride-sharing, micro-transit, and courier network services. Several services allow their users to share different types of mobility, including cars and micromobility. Furthermore, with the advancement of automated vehicle (AV) technology, shared mobility services may provide important alternatives to conventional transportation and have the potential to alter the way in which people move around cities. A convergence of these two innovations is beginning to develop, with small-scale shared automated vehicle (SAV) tests emerging around the world \cite{Stocker2017Shared}. In fact, there has been much speculation regarding the effects of shared automated mobility on traveler behavior, traffic congestion, and the environment \cite{Stocker2017Shared}.

However, as shared automated vehicles (SAVs) are popularized, it is increasingly important to study how users trust such systems \cite{hartl_sustainability_2018,mittendorf_implications_2017}, considering that not all services provide full autonomy (SAE Level 4 or 5) and instead several remain at partial (SAE Level 2) automation. Additionally, one of the most challenging, yet essential aspects of future SAV adoption is calibrating users' trust in the service and technology, particularly if they are only using SAVs on an `as-needed' (i.e., not daily) basis. It is therefore incumbent upon researchers to study users' trust in different types of automated mobilities and how their trust may change when switching from one automated mobility type to another.

\subsection{Establishing trust in automation across shared mobility}

It is widely accepted that successful adoption of automated technologies relies on user trust \cite{lee_trust_2004}. Adoption of automation is dynamic and influenced by individual perceptions, attitudes, and beliefs that rely on effective human–technology coagency \cite{ghazizadeh2012extending}.

To evaluate a humans' trust in automation, examining interpersonal trust towards other humans may provide insight on their trust in automation. \citeauthor{rempel_trust_1985} \cite{rempel_trust_1985} proposed that interpersonal trust in humans evolves from a state of dependability, predictability, and finally a state of faith. \citeauthor{korber_theoretical_2018} \cite{korber_theoretical_2018}
proposed a framework to bridge the gap between interpersonal trust and trust in automation using the theoretical models proposed by \citeauthor{mayer_integrative_1995} \cite{mayer_integrative_1995} and \citeauthor{lee_trust_2004} \cite{lee_trust_2004}, respectively. In this paradigm, they proposed that trust depends on system reliability/competence, understandability/predictability, intention of developers, familiarity, perceived trustworthiness, and general trust in automation \cite{korber_theoretical_2018}. This framework allows researchers to establish the relationship between dispositional trust amongst humans and general trust in automation systems, while trying to establish the dynamics of trust. 

\subsection{Shared understanding between drivers and automation in mobility}

Designing automated systems for aggressive drivers is of particular interest to the research community. Aggressive drivers have had the tendency to display impatience, frustration, and annoyance with other drivers, and tend to display impatience, frustration, and annoyance with other drivers \cite{li2004reliability}\cite{beck2006concerns}. The research found aggressive drivers are more likely to speed, drive with an impairment, have less regard for pedestrian safety and disabled people \citeauthor{tasca2000review}. \cite{tasca2000review} in their review of aggressive driving discussed that the behaviors reported across drivers could engage in deliberate dangerous driving (purposeful) which could lead to risk for all types of road users. With this establishing aggressive driving behavior as an inherent behavioral trait, it is interesting to consider how aggressive drivers would accept automated vehicles that are likely to exhibit behavior that matches their preferred driving style, and does that lead to acceptance of different automated mobility. A particular challenge remains for aggressive drivers who may prefer their mobility to adopt a driving trajectory with higher speed, small headway and gap, higher accelerations \cite{sagberg2015review}. Past research has established that aggressive driver may not accept automated vehicles with defensive driving styles. \citeauthor{ma_investigating_2020} \cite{ma_investigating_2020} explored the impact of driver’s driving styles and AV’s driving styles on trust, acceptance, and takeover behaviors. Additionally, \citeauthor{bellem_comfort_2018} \cite{bellem_comfort_2018} explored how introducing traits of manual driving (jerkiness while accelerating) could lead to better comfort for passive drivers in an automated vehicle. 

\subsection{Objectives of the study}
As discussed in the previous section, there is a dearth of literature on how trust  changes between two different automated mobilities. The objectives of this research are to identify how multiple dimensions of trust vary between automated mobilities, and whether trust builds the same or differently as users transition between different mobilities. Additionally, it is important to consider how different dimensions of trust are affected when an individual uses another mobility after a (time) gap in interaction. 
The two key questions that are explored in this paper are: 
\begin{enumerate}
    \item Does trust build similarly across different mobility types?
    \item  Does trust transfer differently across different mobility types?
\end{enumerate}
In the next section, the details of the experimental study and the experimental protocol are explained in further detail.

\section{Methods}

\subsection{Participants}

Forty-eight participants (28 Male and 20 Female) were recruited from the greater San Jose area in California, USA to participate in the study. The age of the participants ranged between 19 and 69 years, with an average age of 32.83 years (standard deviation of 11.53 years). The participants completed the written consent for participating in the study. The study was approved by the Institutional Review Board at San Jose State University. All participants were required to: 1) be legally allowed to drive in the United States; 2) be older than 18 years of age; 3) have no self-reported hearing impairment; 4) have perfect or corrected vision using contact lenses (glasses could not be accommodated in the virtual reality headset); and 5) not easily susceptible to motion sickness. Participants were compensated \$125 for their time. Participants who did not complete the entire experiment were compensated \$25.  

\subsection{Experimental Scenarios}
\begin{figure}[!ht]
  \centering
  \includegraphics[width=0.75\textwidth]{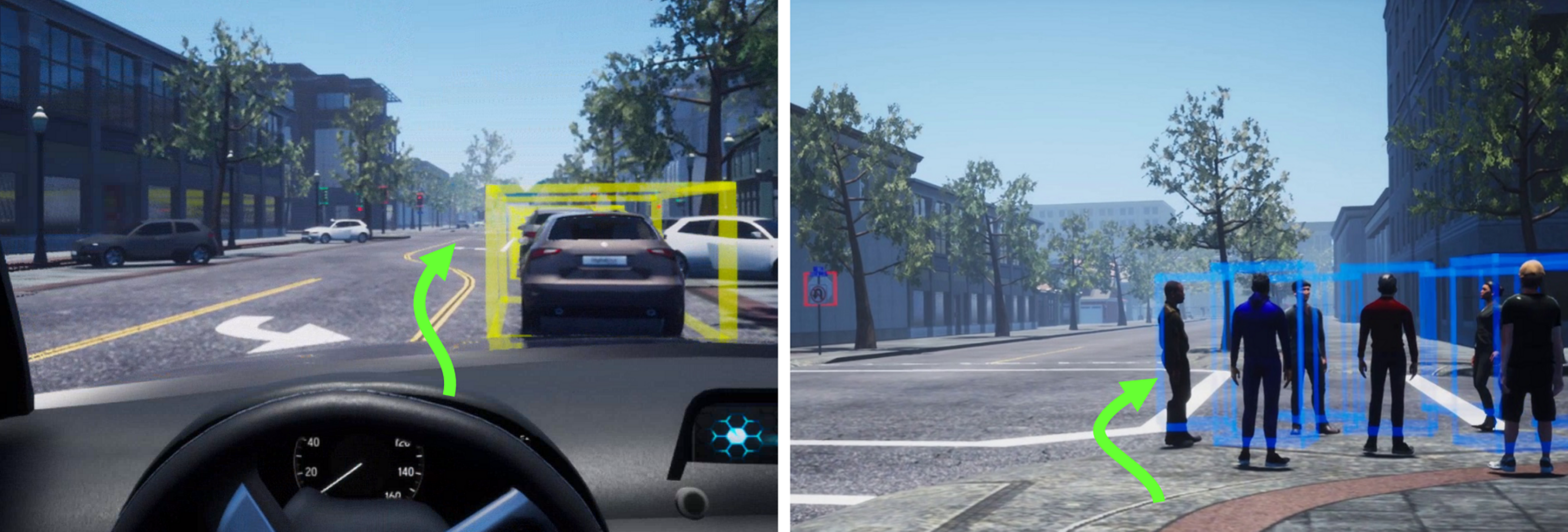}
  \caption{Screenshots showing example proactive car event and sidewalk event, respectively. The green arrow shows the future path that will be traversed. Left image shows the car event ``Crossing Traffic Lines (double yellow)'' where the car crosses the double yellow line in order to get into the turn lane when there is stopped traffic ahead. Right image shows the sidewalk mobility event ``Crossing Traffic Lines (crosswalk lines)'' where the sidewalk mobility drives onto the road outside the crosswalk lines because of stopped pedestrians ahead.}\label{fig:scenario_example}
\end{figure}

The experiment consists of a variety of automated vehicle drives aimed at eliciting a trust response from the participant. These include a tutorial drive, a standard drive, and a ``proactive'' drive, all described below. To manipulate trust without changing automation reliability, a set of events was created in which the automation performed ``proactive'' maneuvers while maintaining safe driving behavior. The goal of these events was to create situations in which the automated vehicle had multiple options of safe actions to perform. For each event, two actions were designed; one represented an ``aggressive'' action, while the other represented a ``defensive'' action. Each proactive drive had 8 events, as listed in Table~\ref{tab:proactive_events}. For example, in the case of the ``stale green light'' event, the aggressive automated vehicle would accelerate to cross the traffic light before it turned yellow or red, whereas the defensive automated vehicle would slow down, assuming that it did not have enough time to cross the traffic light. Another example proactive event is shown in Figure~\ref{fig:scenario_example}. Since all car events were not directly applicable for the sidewalk mobility, equivalent events were created to match the car events as closely as possible. Standard drives involved no proactive events in order to serve as prerequisite trust-building drives (prior to proactive drives). They involved the automated vehicle navigating through multiple intersections in an urban area. Similar to the proactive drives, standard drives included the presence of other cars and pedestrians throughout the urban area; however, there were no ambiguous scenarios which required advanced decision-making by the automated vehicle. Each drive lasted for approximately 8 minutes. Finally, the tutorial consisted of a simple drive through an empty urban area with no other cars or pedestrians, which lasted for approximately 3 minutes.

\begin{table}[!ht]
\small
	\caption{List of proactive events}\label{tab:proactive_events}
	\begin{center}
		\begin{tabular}{p{0.4\linewidth} | p{0.43\linewidth}}\toprule
			Car Events & Sidewalk Mobility events \\\midrule
			Yellow light & Yellow light\\
            Left turn yield (green light) & Yield to turning car\\
            Jay walking pedestrians & Jay walking pedestrians\\
            Crossing Traffic Lines (double yellow) & Crossing Traffic Lines (crosswalk lines)\\
            Right turn merge (red light) & Merge into crowd of pedestrians\\
			Simultaneous arrival at intersection & Simultaneous arrival at intersection\\
            Car sudden backout from driveway & Pedestrian runout from behind house\\
            Passing slowing car & Passing slow pedestrians\\\bottomrule
		\end{tabular}
	\end{center}
\end{table}

To ensure that each participant remained aware of all the actions of the automated mobility, particularly during proactive scenarios, two different forms of transparency were added. First, a navigation arrow was added to mimic a GPS navigation system and was used to indicate the path of the automated vehicle. Next, a text-to-speech audio cue was played at the onset of each proactive event. The text was tailored for each specific event and differed depending on whether the event was defensive or aggressive and whether the vehicle was a car or sidewalk mobility. Aggressive audio cues are listed in Table~\ref{tab:audio_cues_agg}.
	
\begin{table}[!ht]
\small
	\caption{Text-to-Speech Audio Cues for Aggressive Car and Sidewalk Mobility Drives}\label{tab:audio_cues_agg}
	\begin{center}
		\begin{tabular}{p{0.24\linewidth} | p{0.35\linewidth} | p{0.35\linewidth}}\toprule
		Event & Car Audio Cues & Sidewalk Mobility Audio Cues \\\midrule
		Stale green & Yellow light ahead; maintaining speed & Yellow light ahead; maintaining speed\\
		Yield scenario & Heavy traffic ahead; turning at gap & Crowd ahead; continuing at gap\\
		Jaywalking pedestrians & Pedestrians stopped; continuing & Pedestrians stopped; continuing\\
		Crossing traffic lines & Left turn lane blocked; crossing lines to pass & Sidewalk blocked; crossing lines to pass\\
		Merge into traffic & Heavy traffic ahead; turning at gap & Crowd ahead; turning at gap\\
		Simultaneous arrival & No audio cue & No audio cue\\
		Sudden back out & Car ahead stopped; continuing & Pedestrian stopped; continuing\\
		Passing slow traffic & Car ahead slowed to a stop; passing & Slow pedestrian ahead; passing\\\bottomrule
		\end{tabular}
	\end{center}
\end{table}

\subsection{Design of experiment}\label{sec-expdesign}
To address the research objectives, an experiment was designed to explore how participants' trust transferred across different automated mobilities. Participants were assigned to the automation driving style that they preferred to be driven in (as opposed to their personal driving style), based upon a pre-experiment survey described in the next section. This was done to avoid confounding effects of perceived capability and driving style mismatch on participants' trust. Out of 48 participants recruited for the study, 24 participants were assigned to the aggressive driving style of automation and the remaining 24 participants were assigned to the defensive driving style. The difference between the driving styles was strictly posited around gap acceptance, smoother acceleration and deceleration, and driving over and under speed limits. 

\begin{figure}[!ht]
  \centering
  \includegraphics[width=0.99\textwidth]{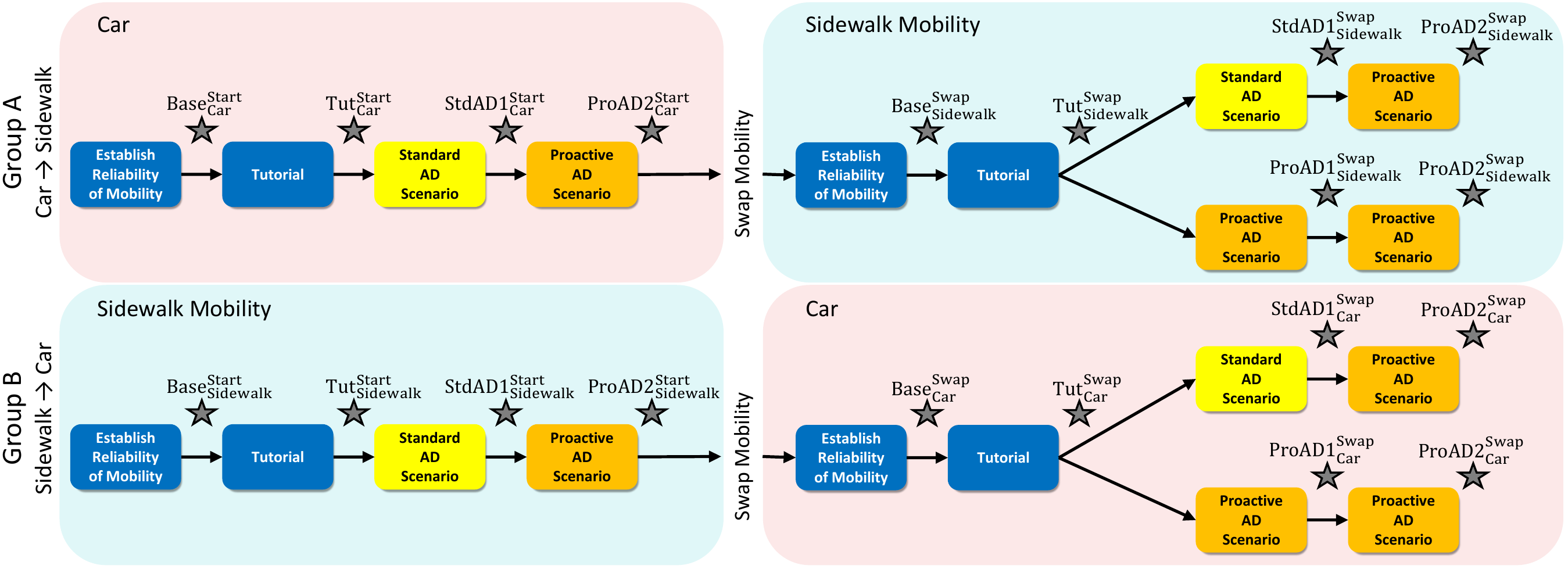}
  \caption{Experiment design illustration.}\label{fig:expdesing}
\end{figure}

Once the participants were assigned to their preferred style, they were assigned to one of two mobility types: car or sidewalk. Half of the participants interacted with the car mobility first, while the other half interacted with the sidewalk mobility first. To eliminate the effect of perceived capability of the automation, the reliability of the mobility was established at the beginning of the experiment. The participants are informed that ``The hardware and software of the vehicle is 100\% reliable. The vehicle is brand new, so all the hardware has been thoroughly tested and meets industry standards. The self-driving software has been tested over millions of miles and is suited for consumer use in everyday driving.''  In their first mobility, once the reliability of the mobility was established, each participant completed a ``tutorial'' drive followed by a ``standard'' and then a ``proactive'' drive. After completing 3 drives on the first mobility, participants changed mobility. Half of the participants received a ``standard'' then ``proactive'' drive on the second mobility, while the other half received a ``proactive'' drive followed by another ``proactive'' drive, after establishing reliability of mobility and the tutorial drive. The swapping of different mobility types allowed for the observation of trust transfer between car to sidewalk mobility and sidewalk mobility to car, providing information on mobility order. Additionally, dividing participants to compare ``standard $\rightarrow$ proactive'' versus ``proactive $\rightarrow$ proactive'' created the opportunity to examine the effective drive-type within a given mobility after a swap. The workflow of the experiment design is shown in Figure~\ref{fig:expdesing}.

\subsection{Measuring trust}
Trust was measured four times per mobility interaction, as shown by the gray stars in Figure~\ref{fig:expdesing}. Each gray star is labeled  based on the drive, mobility, and whether the drive was in the start of or after the mobility swap. For example, $Tut^{Swap}_{Car}$ means that the survey responses were collected after the tutorial drive for the car mobility after the mobility was swapped (i.e., swapped from sidewalk to car). In order to capture different aspects of trust pertinent to the research interests of this study, we adapted a trust in automation questionnaire developed by \citeauthor{korber_theoretical_2018} in 2018 \cite{korber_theoretical_2018}. Although \citeauthor{korber_theoretical_2018}’s trust questionnaire originally contained six dimensions of trust, we only used five dimensions and excluded ``familiarity''. In their theoretical model, familiarity itself is not considered to be an element of trust in automation, but indirectly influences it as a moderator. As recommended by \citeauthor{korber_theoretical_2018}, we eliminated the ``familiarity'' focus on the ``core questionnaire''. The adapted trust survey questions and their respective dimensions are shown in Table~\ref{tab:trust_survey}.

\begin{table}[!ht]
\small
	\caption{Trust in Automation Survey (adapted from \cite{korber_theoretical_2018})} \label{tab:trust_survey}
	\begin{center}
		\begin{tabular}{p{0.66\linewidth} | p{0.27\linewidth}}\toprule
			Item & Scale\\\midrule
		    The vehicle is capable of interpreting situations correctly & Reliability/Competence\\
            The vehicle works reliably & Reliability/Competence\\
            A vehicle malfunction is likely & Reliability/Competence\\
            The vehicle is capable of taking over complicated tasks & Reliability/Competence\\
            The vehicle might make sporadic errors & Reliability/Competence\\
            I am confident in the vehicle’s self-driving capabilities & Reliability/Competence\\
            The vehicle’s state is always clear to me & Understanding/Predictability\\
            The vehicle acts unpredictably & Understanding/Predictability\\
            I am able to understand why the vehicle responds in certain ways &  Understanding/Predictability\\
            It’s difficult to identify what the vehicle will do next &  Understanding/Predictability\\
            The vehicle’s developers are trustworthy & Intention of Developers\\
            The vehicle’s developers take my well-being seriously & Intention of Developers\\
            I should be careful with the self-driving vehicle & Propensity to Trust\\
            I’m prone to trusting the self-driving vehicle than mistrusting it & Propensity to Trust\\
            The self-driving vehicle generally works well & Propensity to Trust\\
            I trust the vehicle & Trust in Automation\\
            I can rely on the vehicle & Trust in Automation\\\bottomrule
		\end{tabular}
	\end{center}
\end{table}

\subsection{Hardware}
\subsubsection{Headset}
A StarVR headset with 210-degree field-of-view (FoV) was used for both mobilities in this experiment. Several actors in the simulated scene were programmed to create some of the proactive events; therefore, in order to improve participants’ situational awareness (SA), a VR headset with a broad FoV was chosen. Additionally, the StarVR headset has built-in eye tracking capabilities, which allowed us to capture where people were looking at any given moment.

\subsubsection{Motion Base and Vehicle Platforms}
\begin{figure}[!ht]
  \centering
  \begin{subfigure}[b]{0.4\textwidth}
     \centering
     \includegraphics[width=\textwidth]{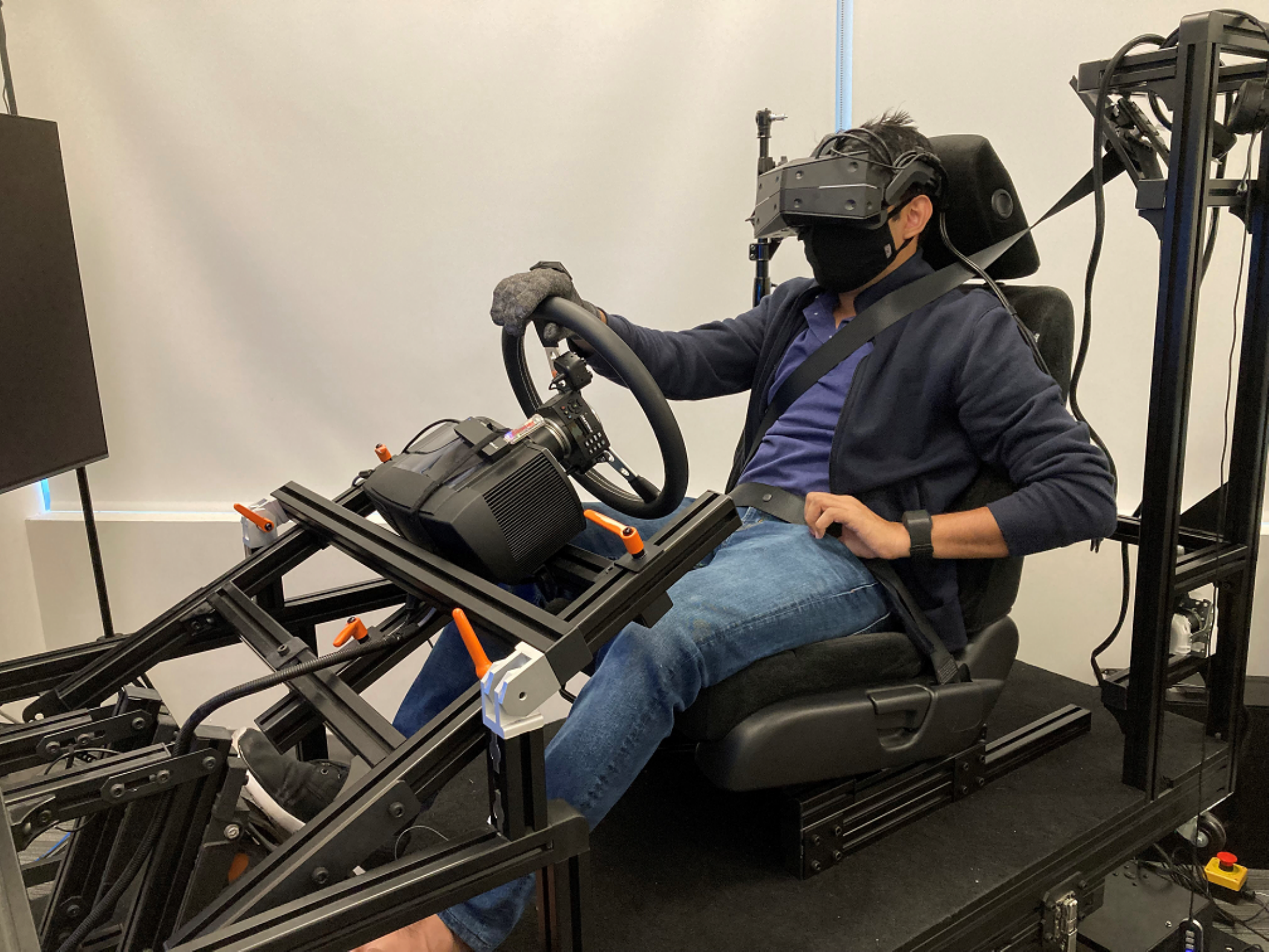}
     \caption{Car mobility platform}
     \label{fig:trial_car}
  \end{subfigure} \qquad
  \begin{subfigure}[b]{0.4\textwidth}
     \centering
     \includegraphics[width=\textwidth]{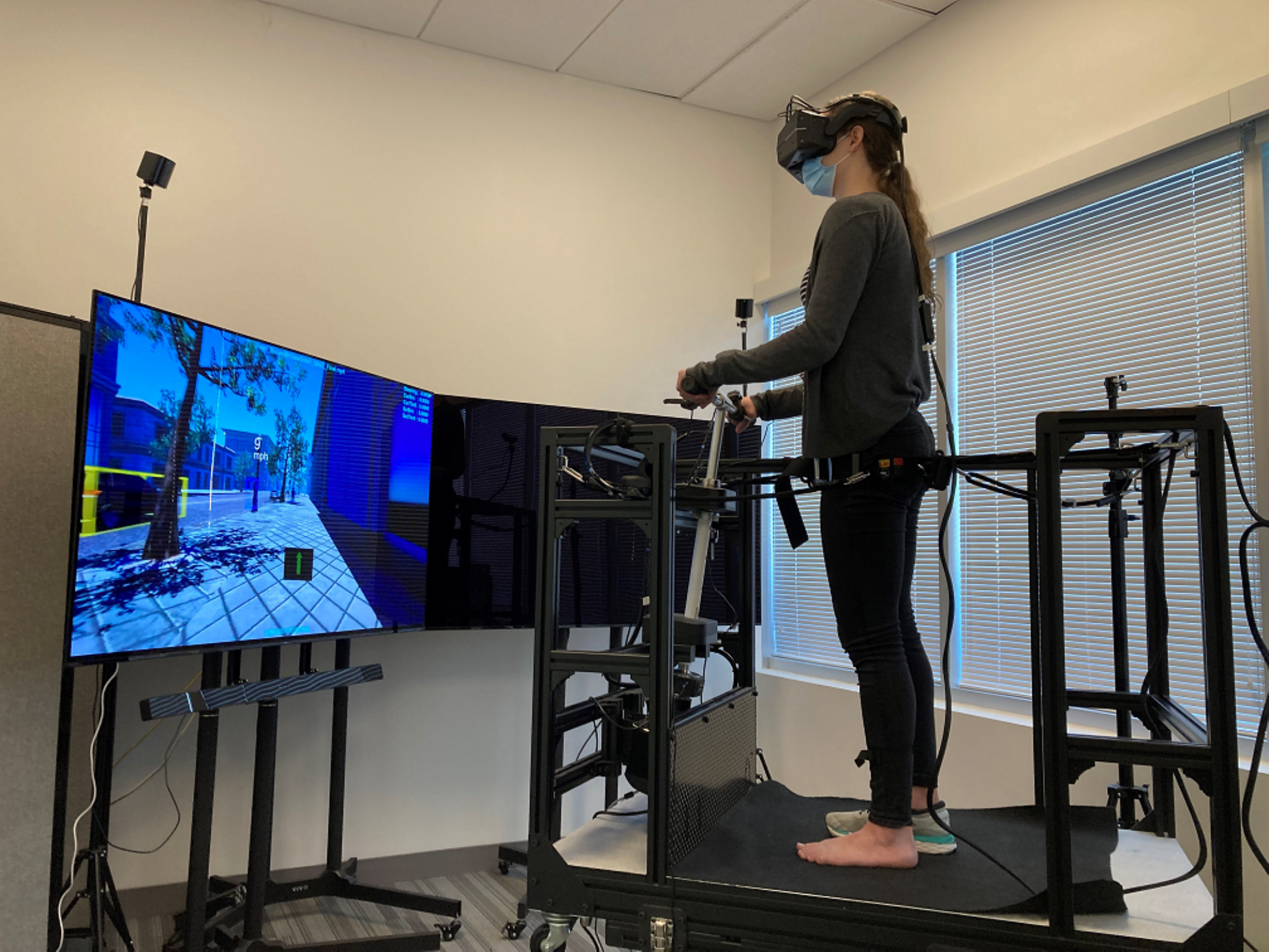}
     \caption{Sidewalk mobility platform}
     \label{fig:trial_sm}
  \end{subfigure} 
  \caption{Mobility platforms mounted on the motion base.}\label{fig:trial}
\end{figure}

A motion base allowed for rotational movement in three degrees of freedom: pitch, yaw, and roll. This allowed for a higher fidelity driving simulation in which participants could feel typical forces experienced in or on a vehicle. Two custom vehicle platforms compatible with the motion base were used throughout the experiment. The first was a car platform and the second was a sidewalk mobility platform (see Figure~\ref{fig:trial}).

\subsubsection{Dimensions for analysis}

For this study, the dependent variables were different dimensions of trust as defined by the survey questions shown in Table~\ref{tab:trust_survey}. All the self-reported trust survey responses were normalized for each individual participant by using z-normalization, except for the results presented in ``Variation in trust across different drive type.'' This ensured that the analysis accounted for any dispositional trust that a participant brought with them to the experiment when analyzing trust across mobilities and within participants. The trust dimensions were compared across the different types of drives and mobility types. In the next section, the findings from the analysis are presented.

\section{Results}

In this section, we first compare the evolution of trust over different drives and across mobilities. Next, a comparison between baseline trust as well as the reported trust post-tutorial for each mobility is conducted. Finally, the impacts of swapping mobilities on different dimensions of trust are reported. 

\subsection{Variation in trust across different drive type}

\begin{figure}[!ht]
  \centering
  \begin{subfigure}[b]{0.48\textwidth}
     \centering
     \includegraphics[width=\textwidth]{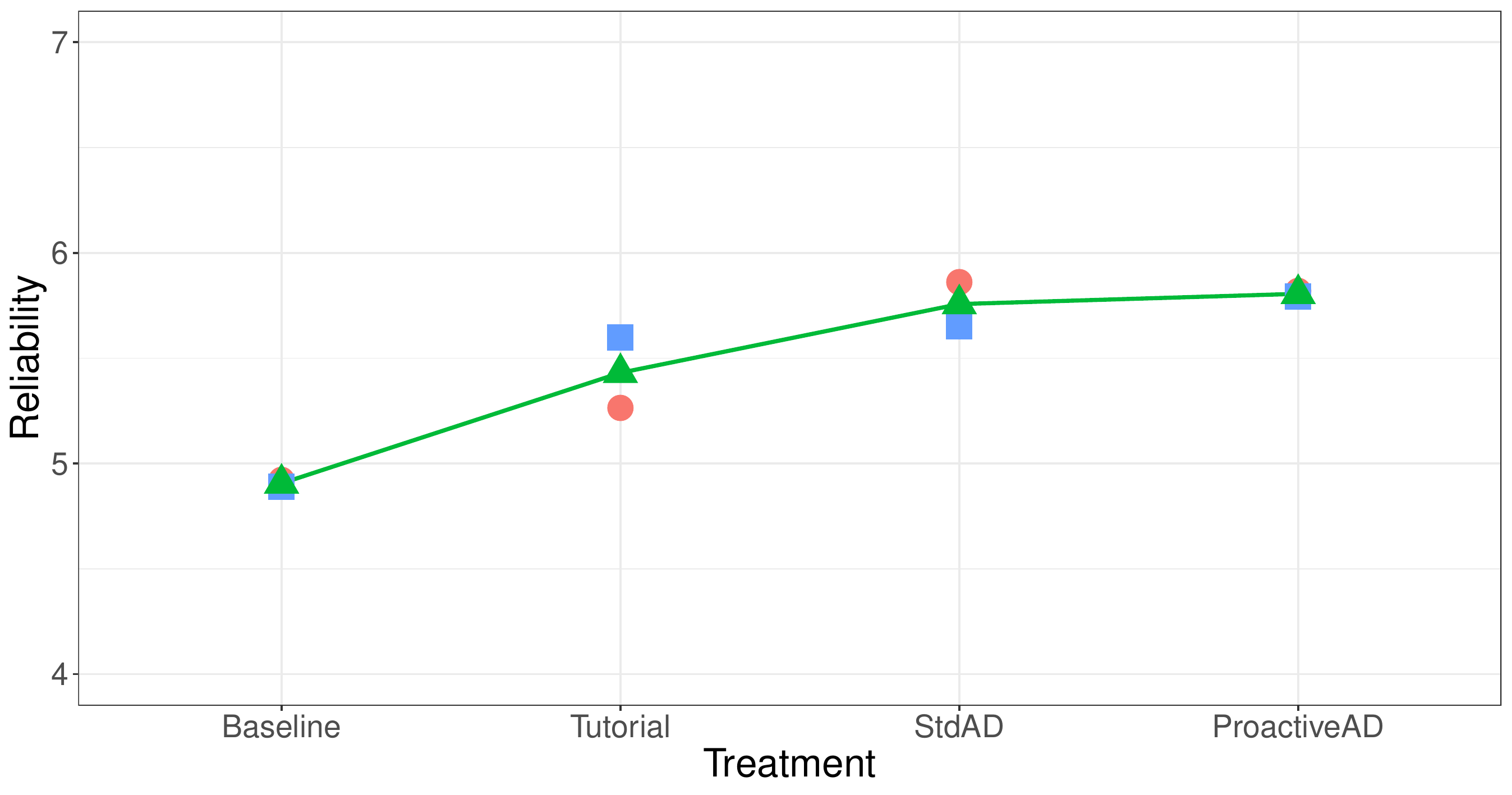}
     \caption{Reliability}
     \label{fig:t_reliability}
  \end{subfigure} 
  \begin{subfigure}[b]{0.48\textwidth}
     \centering
     \includegraphics[width=\textwidth]{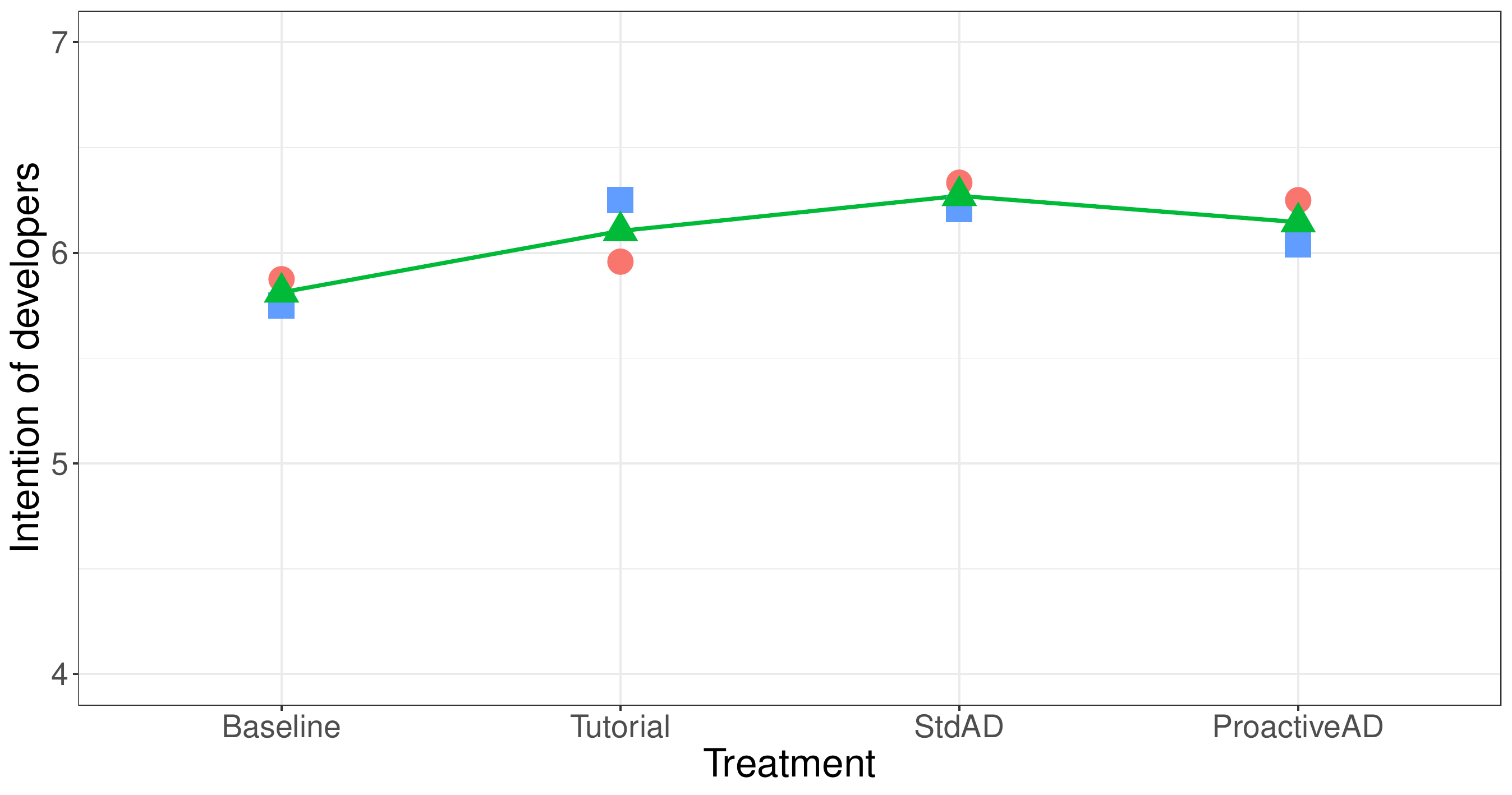}
     \caption{Understanding}
     \label{fig:t_understanding}
  \end{subfigure} \\
  \begin{subfigure}[b]{0.48\textwidth}
     \centering
     \includegraphics[width=\textwidth]{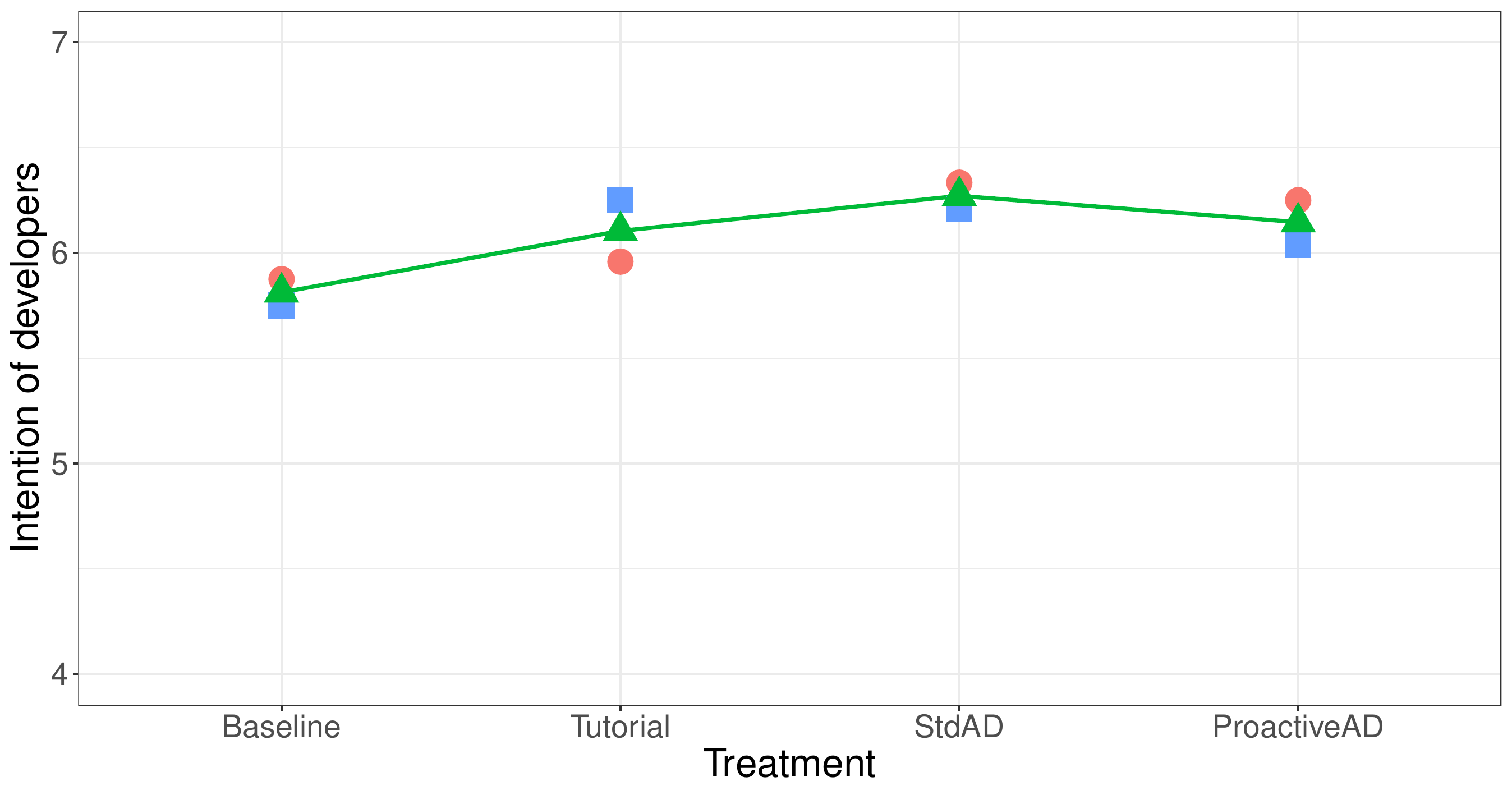}
     \caption{Intention of developers}
     \label{fig:t_intention}
  \end{subfigure} 
  \begin{subfigure}[b]{0.48\textwidth}
     \centering
     \includegraphics[width=\textwidth]{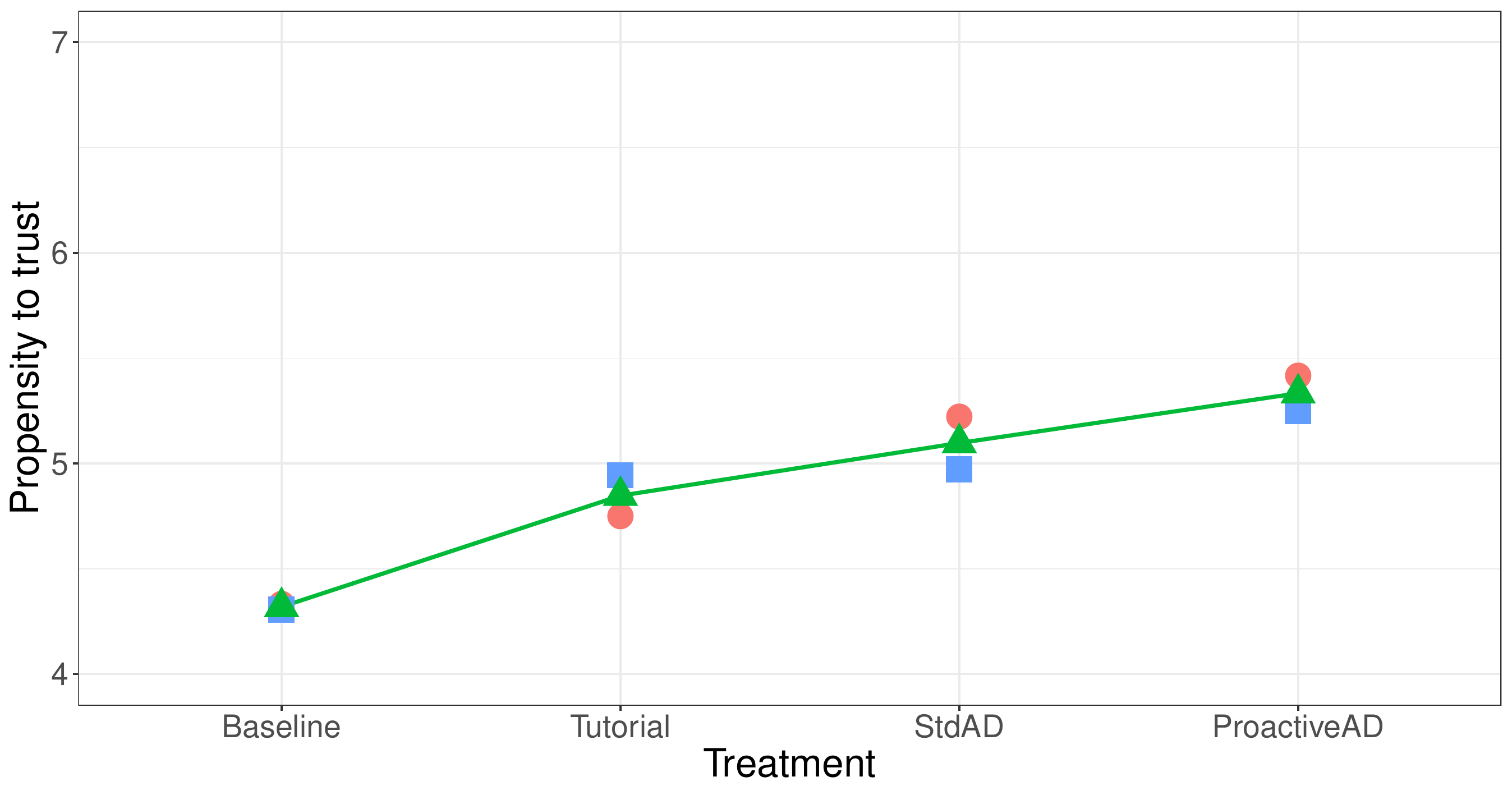}
     \caption{Propensity to trust}
     \label{fig:t_propensity}
  \end{subfigure} \\
  \begin{subfigure}[b]{0.48\textwidth}
     \centering
     \includegraphics[width=\textwidth]{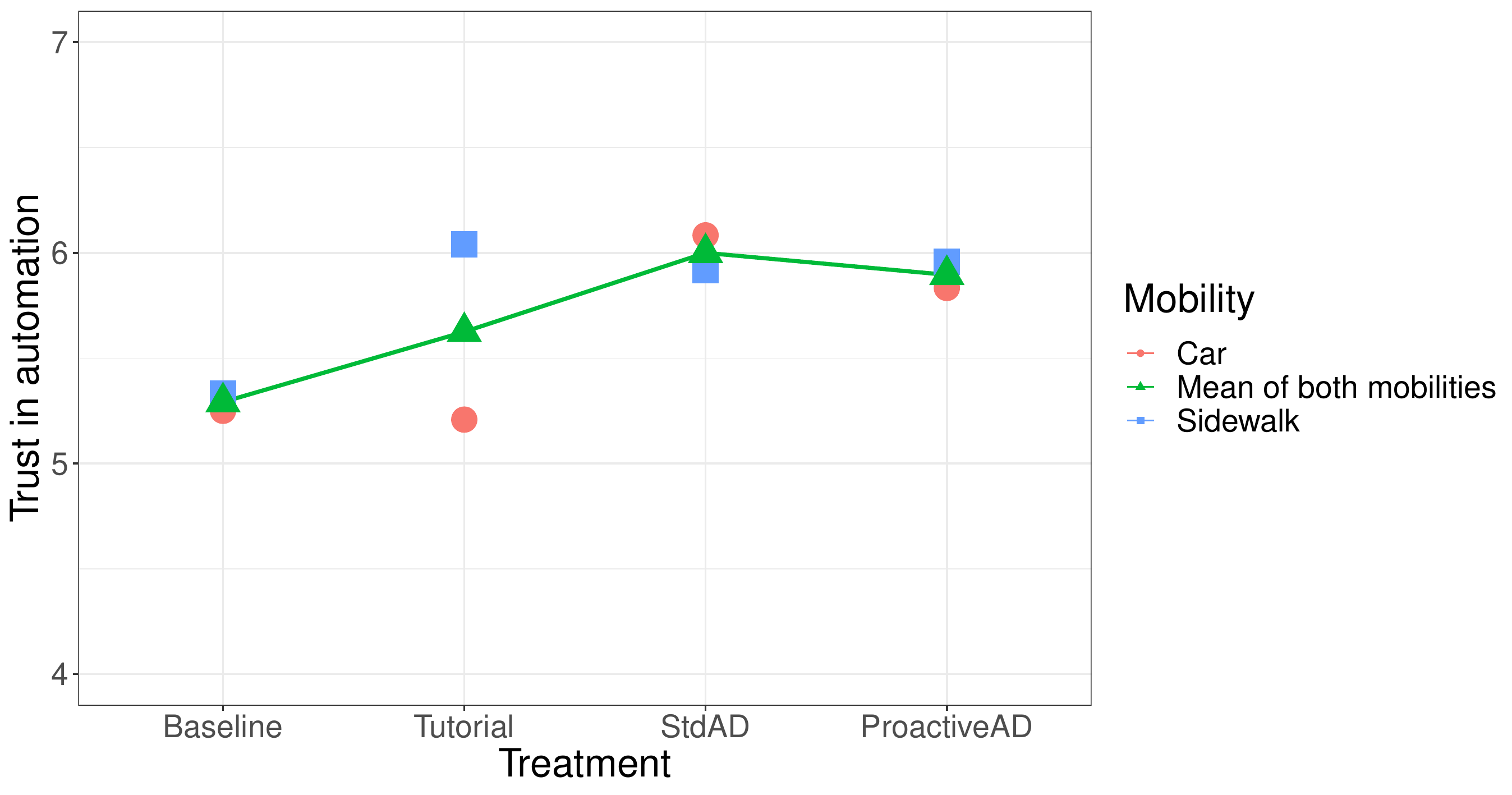}
     \caption{Trust in automation}
     \label{fig:t_trust}
  \end{subfigure} 
  \caption{Comparison of the mean values of trust dimensions when considering the initial mobility (Start Mobility).}\label{fig:t}
\end{figure}

We found that there is a significant effect of drive type during participants' initial mobility interaction on all five dimensions of trust:

\begin{itemize}
    \item Reliability/Competence: \emph{F}(3,69) = 18.57, p < .00001, $\eta_{p}^2 = 0.45$
    \item Understanding/Predictability: \emph{F}(3,69) = 9.78, p < .0001, $\eta_{p}^2 = 0.30$
    \item Intention of Developers: \emph{F}(3,69) = 2.94, p < .05, $\eta_{p}^2 = 0.11$
    \item Propensity to Trust: \emph{F}(3,69) = 9.43, p < .0001, $\eta_{p}^2 = 0.29$
    \item Trust in Automation: \emph{F}(3,69) = 4.25, p < .01, $\eta_{p}^2 =0.16$
\end{itemize}
Additionally, we found that there is no significant effect of mobility type on any of the dimensions of trust. Therefore, there is an overall consistent trend across all five dimensions of trust and across mobilities. The trends across all dimensions show that there is an increase in the mean value post tutorial in the sidewalk mobility as shown in Figure~\ref{fig:t}, which is indicative of an increased acceptance of the mobility post tutorial. However, for the other drive types, the car mobility seems to have a higher mean value as compared to sidewalk mobility. While the figure does show a slightly increasing trend, the difference between the mean values for different dimensions were not statistically significant.

\subsection{Baseline trust dimensions across mobility}
For these comparisons, a pairwise t-test was used to compare the responses across the baseline drive for each group of participants \cite{carmer1985pairwise}. The \textit{t}-score and \textit{p}-values for the statistically significant comparisons are shown in Table~\ref{tab:baseline}.The analysis found that participants who first experienced the car mobility and then switched to the sidewalk mobility, or the other way around did not find any statistical differences between the two baseline states.

\begin{table}[!ht]
\small
\centering
\caption{t-scores for baseline trust dimensions across mobilities.}
\label{tab:baseline}
\begin{tabular}{@{}llrrrr@{}}
\toprule
 &                          & $M$      & $SD$     & $t$ & $p$ \\ \midrule
Reliability & $Base_{Start}^{Car}$ & $-0.8579$            & $0.8145$             & \multirow{2}{*}{$-0.511$} & \multirow{2}{*}{$0.6198$} \\[0.45em]
\quad$Car \rightarrow Sidewalk$ & $Base_{Swap}^{Sidewalk}$ & $-0.5955$ & $1.1142$ &     &     \\ \midrule
Reliability & $Base_{Start}^{Sidewalk}$ & $-0.7808$            & $1.0817$             & \multirow{2}{*}{$-0.291$} & \multirow{2}{*}{$0.7763$} \\[0.45em]
\quad$Sidewalk \rightarrow Car$ & $Base_{Swap}^{Car}$ & $-0.6116$ & $1.1289$ &     &     \\ \midrule
Understanding & $Base_{Start}^{Car}$ & $-0.7753$            & $0.9980$             & \multirow{2}{*}{$-0.481$} & \multirow{2}{*}{$0.6399$} \\[0.45em]
\quad$Car \rightarrow Sidewalk$ & $Base_{Swap}^{Sidewalk}$ & $-0.5556$ & $1.0186$ &     &     \\ \midrule
Understanding & $Base_{Start}^{Sidewalk}$ & $-0.7232$            & $0.9687$             & \multirow{2}{*}{$-0.266$} & \multirow{2}{*}{$0.795$} \\[0.45em]
\quad$Sidewalk \rightarrow Car$ & $Base_{Swap}^{Car}$ & $-0.5896$ & $0.9943$ &     &     \\ \midrule
Intention of developers & $Base_{Start}^{Car}$ & $-0.4783$            & $0.8968$             & \multirow{2}{*}{$-0.132$} & \multirow{2}{*}{$0.8974$} \\[0.45em]
\quad$Car \rightarrow Sidewalk$  & $Base_{Swap}^{Sidewalk}$ & $-0.3962$ & $1.3015$ &     &     \\ \midrule
Intention of developers & $Base_{Start}^{Sidewalk}$ & $-0.3555$            & $1.0823$             & \multirow{2}{*}{$0.4144$} & \multirow{2}{*}{$0.6866$} \\[0.45em]
\quad$Sidewalk \rightarrow Car$ & $Base_{Swap}^{Car}$ & $-0.5962$ & $1.1322$ &     &     \\ \midrule
Propensity to Trust & $Base_{Start}^{Car}$ & $-0.8054$            & $0.5576$             & \multirow{2}{*}{$-0.465$} & \multirow{2}{*}{$0.6507$} \\[0.45em]
\quad$Car \rightarrow Sidewalk$  & $Base_{Swap}^{Sidewalk}$ & $-0.6631$ & $0.7300$ &     &     \\ \midrule
Propensity to Trust & $Base_{Start}^{Sidewalk}$ & $-0.4789$            & $1.1813$             & \multirow{2}{*}{$-0.182$} & \multirow{2}{*}{$0.8592$} \\[0.45em]
\quad$Sidewalk \rightarrow Car$ & $Base_{Swap}^{Car}$ & $-0.5861$ & $0.8457$ &     &     \\ \bottomrule
Trust in automation & $Base_{Start}^{Car}$ & $-0.8751$            & $0.6521$             & \multirow{2}{*}{$-0.182$} & \multirow{2}{*}{$0.8592$} \\[0.45em]
\quad$Car \rightarrow Sidewalk$  & $Base_{Swap}^{Sidewalk}$ & $-0.7941$ & $1.0876$ &     &     \\ \midrule
Trust in automation & $Base_{Start}^{Sidewalk}$ & $-0.2694$            & $1.0624$             & \multirow{2}{*}{$0.8025$} & \multirow{2}{*}{$0.4392$} \\[0.45em]
\quad$Sidewalk \rightarrow Car$ & $Base_{Swap}^{Car}$ & $-0.7019$ & $1.0784$ &     &     \\ \bottomrule
\end{tabular}
\end{table}

\subsection{Post tutorial trust dimensions across mobility}
\begin{figure}[!ht]
  \centering
  \begin{subfigure}[b]{0.48\textwidth}
     \centering
     \includegraphics[width=\textwidth]{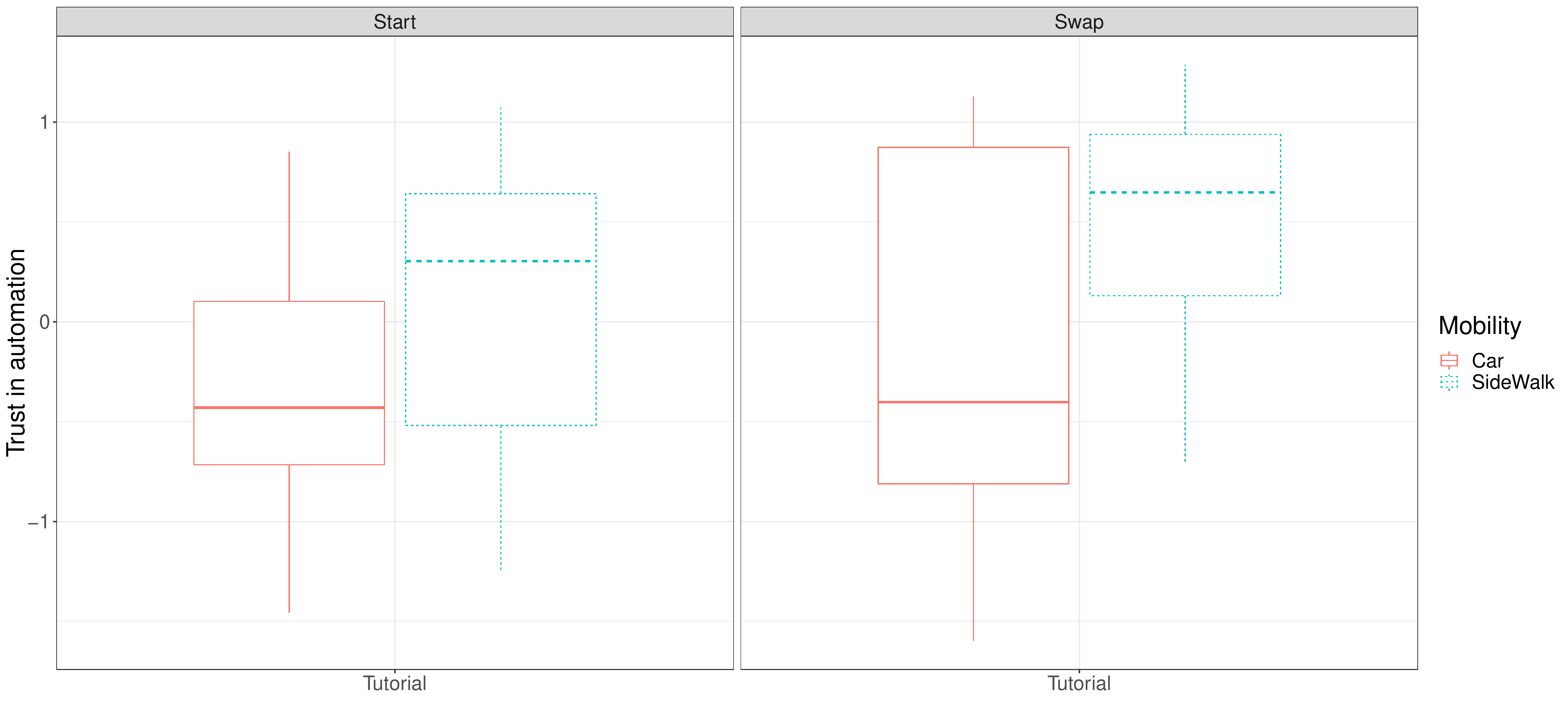}
     \caption{Trust in automation}
     \label{fig:tutorial_tutorial_trust}
  \end{subfigure} 
  \centering
  \begin{subfigure}[b]{0.48\textwidth}
     \centering
     \includegraphics[width=\textwidth]{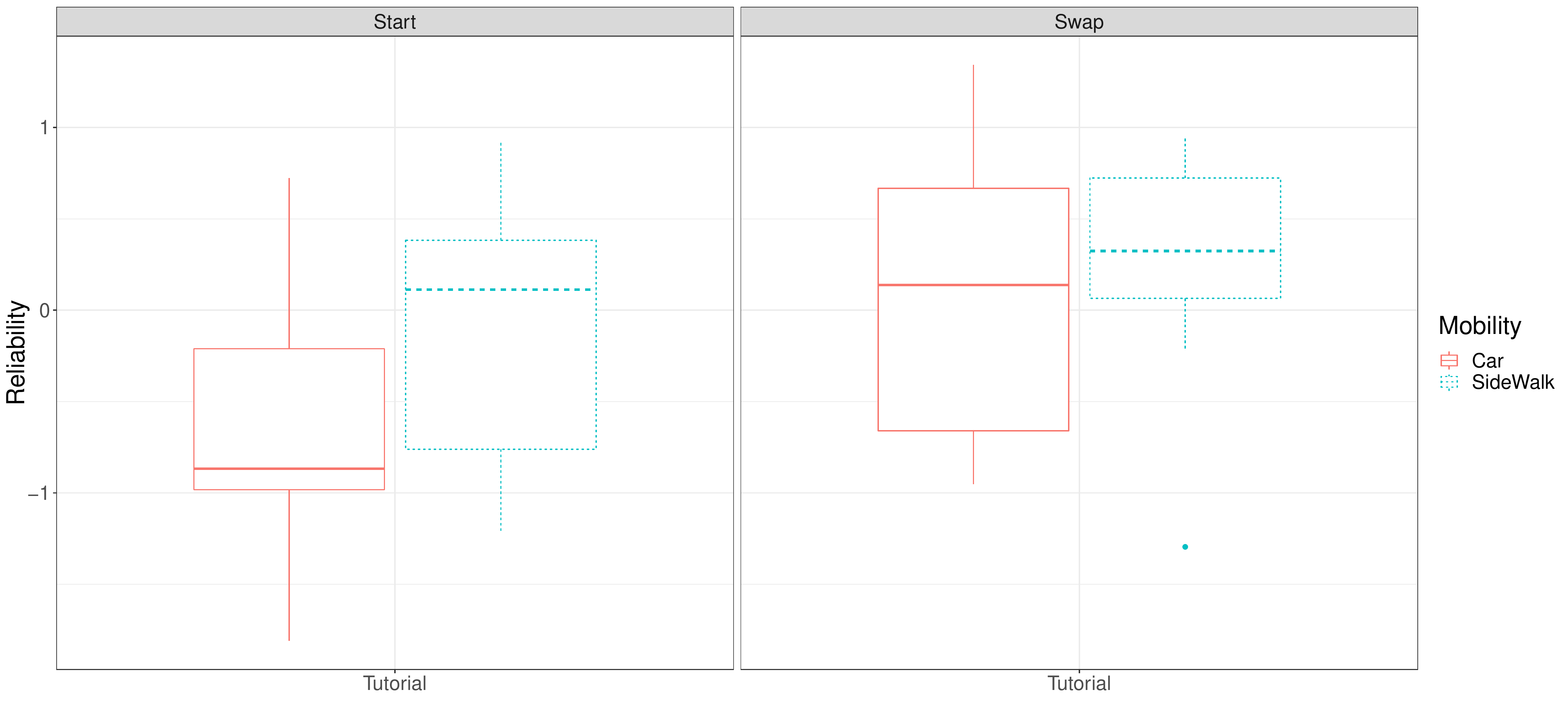}
     \caption{Reliability}
     \label{fig:tutorial_tutorial_reliability}
  \end{subfigure} 
  \caption{Difference in significant trust dimensions post tutorial for each mobility}\label{fig:tutorial_tutorial}
\end{figure}

For these comparisons, a pairwise t-test was used to compare the responses across tutorial drives for each group of participants \cite{carmer1985pairwise}. 
The \textit{t}-score and \textit{p}-values for the statistically significant comparisons are shown in Table~\ref{tab:tutorial}.
\begin{table}[!ht]
\small
\centering
\caption{t-scores for tutorial trust dimensions across mobilities.}
\label{tab:tutorial}
\begin{tabular}{@{}llrrrr@{}}
\toprule
 &                          & $M$      & $SD$     & $t$ & $p$ \\ \midrule
Reliability & $Tut_{Start}^{Car}$ & $-0.6467$            & $0.7807$             & \multirow{2}{*}{$-2.5467$} & \multirow{2}{*}{$0.02715^{***}$} \\[0.45em]
\quad$Car \rightarrow Sidewalk$ & $Tut_{Swap}^{Sidewalk}$ & $0.2551$ & $0.6107$ &     &     \\ \midrule
Reliability & $Tut_{Start}^{Sidewalk}$ & $-0.0859$            & $0.7491$             & \multirow{2}{*}{$-0.3359$} & \multirow{2}{*}{$0.7432$} \\[0.45em]
\quad$Sidewalk \rightarrow Car$ & $Tut_{Swap}^{Car}$ & $0.0565$ & $0.799433$ &     &     \\ \midrule
Understanding & $Tut_{Start}^{Car}$ & $-0.0343$            & $	0.6769$             & \multirow{2}{*}{$-0.57$} & \multirow{2}{*}{$0.5801$} \\[0.45em]
\quad$Car \rightarrow Sidewalk$ & $Tut_{Swap}^{Sidewalk}$ & $0.1683$ & $0.6696$ &     &     \\\midrule
Understanding & $Tut_{Start}^{Sidewalk}$ & $0.0756$            & $0.7057$             & \multirow{2}{*}{$-0.32668$} & \multirow{2}{*}{$0.75$} \\[0.45em]
\quad$Sidewalk \rightarrow Car$ & $Tut_{Swap}^{Car}$ & $0.2048$ & $0.8153$ &     &     \\ \midrule
Intention of developers & $Tut_{Start}^{Car}$ & $-0.3535$            & $0.8319$             & \multirow{2}{*}{$-1.4846$} & \multirow{2}{*}{$0.1657$} \\[0.45em]
\quad$Car \rightarrow Sidewalk$ & $Tut_{Swap}^{Sidewalk}$ & $0.2246$ & $0.7118$ &     &     \\ \midrule
Intention of developers & $Tut_{Start}^{Sidewalk}$ & $0.0197$            & $0.9573$             & \multirow{2}{*}{$-0.09001$} & \multirow{2}{*}{$0.935$} \\[0.45em]
\quad$Sidewalk \rightarrow Car$ & $Tut_{Swap}^{Car}$ & $-0.2954$ & $0.7016$ &     &     \\ \midrule
Propensity to trust & $Tut_{Start}^{Car}$ & $-0.2533$            & $0.6617$             & \multirow{2}{*}{$-1.3751$} & \multirow{2}{*}{$0.1965$} \\[0.45em]
\quad$Car \rightarrow Sidewalk$ & $Tut_{Swap}^{Sidewalk}$ & $0.2286$ & $0.8052$ &     &     \\ \midrule
Propensity to trust & $Tut_{Start}^{Sidewalk}$ & $0.0806$            & $0.9996$             & \multirow{2}{*}{$0.2867$} & \multirow{2}{*}{$0.7797$} \\[0.45em]
\quad$Sidewalk \rightarrow Car$ & $Tut_{Swap}^{Car}$ & $-0.0668$ & $0.8763$ &     &     \\ \midrule
Trust in automation & $Tut_{Start}^{Car}$ & $-0.2906$            & $0.6576$             & \multirow{2}{*}{$-2.3139$} & \multirow{2}{*}{$0.04102^{***}$} \\[0.45em]
\quad$Car \rightarrow Sidewalk$ & $Tut_{Swap}^{Sidewalk}$ & $0.485407$ & $0.6212$ &     &     \\ \bottomrule
Trust in automation & $Tut_{Start}^{Sidewalk}$ & $0.0362$            & $0.7851$             & \multirow{2}{*}{$0.4042$} & \multirow{2}{*}{$0.6938$} \\[0.45em]
\quad$Sidewalk \rightarrow Car$ & $Tut_{Swap}^{Car}$ & $-0.1559$ & $0.9639$ &     &     \\ \bottomrule
\end{tabular}
\end{table}
The results showed participants who first experienced the car mobility and then sidewalk mobility, and had participated in the tutorial, had an increase in \emph{reliability} (see Figure~\ref{fig:tutorial_tutorial_reliability}, $Tut_{Start}^{Car}$ versus $Tut_{Swap}^{Sidewalk}$). The results also indicated that participants who first experienced the car mobility and then the sidewalk mobility, and had participated in the tutorial, had an increase in \emph{Trust in automation} (see Figure~\ref{fig:tutorial_tutorial_trust}, $Tut_{Start}^{Car}$ versus $Tut_{Swap}^{Sidewalk}$). Participants who switched from the car mobility to the sidewalk mobility observed no significant differences in any other dimensions of trust. Additionally, participants who switched from the sidewalk to the car mobility reported no difference in the \emph{trust in automation} dimension of trust, a decrease in \emph{propensity to trust}, and a decrease in the \emph{intention of developers}.

\subsection{\emph{Reliability/competence} dimension of trust development across mobility}
\begin{figure}[!ht]
  \centering
  \includegraphics[width=0.8\textwidth]{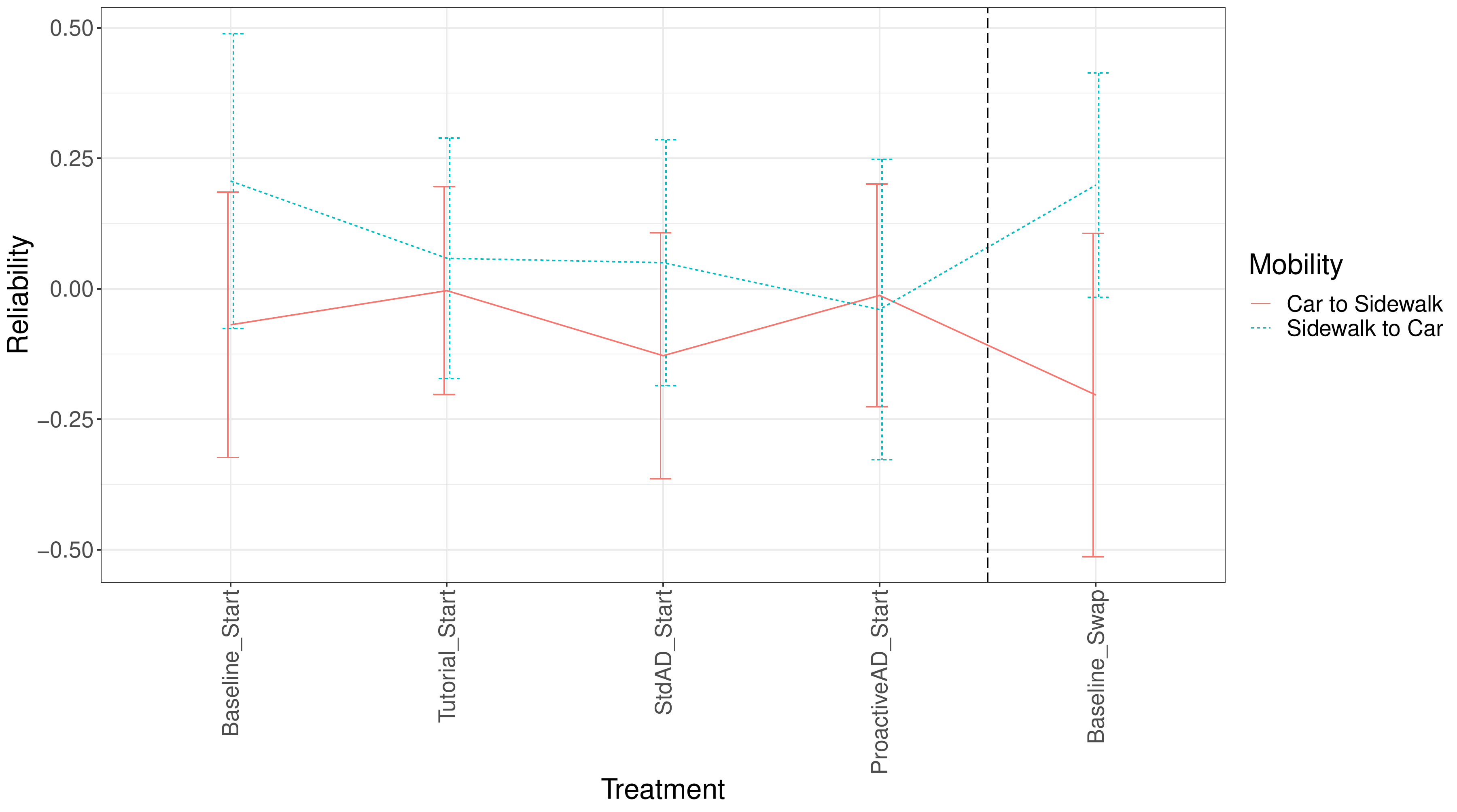}
  \caption{Change in reliability post swap for different mobility types}\label{fig:ig_rplot}
\end{figure}

 Results from the Figure~\ref{fig:ig_rplot} found that trust builds differently depending on the type of mobility transfer. Participants who transitioned from the sidewalk to the car mobility were found to have a significant increase in the reliability, i.e., $ProAD2_{Start}^{Car}$ $\rightarrow$ $Base_{Swap}^{Sidewalk}$: $t (11) = -3.3358$, $p=0.006643$. 

\section{Discussion}

In this section, the implications of the findings are discussed. This research focused only on the data obtained from the aggressive driving condition.
The analysis found that participants who first experienced the car mobility then the sidewalk mobility reported no statistically significant differences across the different dimensions of Trust. These results indicate that initial trust may differ independently, and there is no direct relationship between the baseline trust and the type of interaction a participant may have had previously, or is likely to have with a mobility.

When comparing the participants' responses after the tutorial for each of the two mobilities, participants reported an increase in the following dimensions of trust when they experienced the car first followed by the sidewalk: \emph{reliability},and \emph{trust in automation}. For the other dimensions, the tutorial had an immediate effect on the different dimensions of trust with respect to participants' initial trust. This indicates that participants may have had a strong preference for a given mobility type (car), and tutorials may help improve overall trust, they do not change how trust may transfer across mobilities. The purpose of such analysis is to show whether experience can change a user's trust dynamics; considering the survey responses, it does not seem to be the case.

The second objective of this work was to evaluate how human trust in one automated mobility might transfer to a second, different automated mobility. Results showed that trust builds differently depending on the type of mobility transfer. Specifically, participants who transitioned from car to sidewalk mobility reported a significant increase in the \emph{reliability/competence} dimension of trust after they swapped the mobility; but their reliability dimension of trust increased during their interaction with the car mobility. However, such a trend is not observed for the case when the participants swap to sidewalk mobility from car mobility; the reliability dimension of trust does not significantly change throughout the interaction with the car mobility and does not significantly change after the swap. This indicates that trust possibly transferred when moving from car to sidewalk mobility, however may not have transferred when moving from sidewalk to car. 

The findings of this study indicate that when engaging with different mobility types enabled with automation, the type of mobility does seem to influence how trust may build and transfer when switching between mobilities. It is worth considering that the exposure to different mobilities was limited by the exposure time. Future study designs could incorporate longer exposure to mobilities, which may have a different outcome with respect to trust dynamics. 
Given that the trust survey analyses themselves do not sufficiently explain the cause for such differences in the various trust dimensions, the experiment also included the collection of qualitative responses from the participants through a semi-structured interview. While a rigorous analysis of the qualitative responses is beyond the scope of this study, 
some participants reported trust to a certain degree at the beginning of their first drive and then expressed that they fully trusted the mobility by the end of the experiment. Some participants mentioned growing accustomed to the automation’s behavior over time. However, with respect to the sidewalk mobility, some participants stated that 
they were either not trusting or unsure of the sidewalk mobility’s ability to transport them safely. Some of their reasoning for this opinion was the lack of a physical enclosure when riding on the sidewalk mobility as compared to the physical protection offered by a car. Several participants did not appreciate the unpredictability of pedestrians on the sidewalk, whereas with the car mobility they expressed an acceptance in terms of the expected behavior due to knowing the set of laws or standards for cars on the road. While these qualitative findings are mostly anecdotal, and certainly require further analysis, they do support this investigation and give insights on potential the root causes for differences in trust across different automated mobilities.

The results from this analysis are only limited to aggressive drivers. Defensive drivers may exhibit different trends, which will be analyzed in future work. Additionally next steps include investigation of potential correlations between self-reported trust dimensions with demographics, takeover intent in different scenarios, as well as monitoring the physiological state of the participants.
The key takeaways from this study can be summarized as follows:
\begin{itemize}
\item	The tutorial helped increase the trust of participants in sidewalk automated mobilities.
\item	The baseline trust in participants is unrelated
\item	The findings confirm that the reliability dimension of trust builds differently when transitioning across different mobilities. Specifically, the trust dimension \emph{reliability/competence} increased when the participants transitioned from car to sidewalk mobility.
\end{itemize}

\subsection{Conclusion}
The overall objective of this research was to establish how participants trust different automated mobilities. A dual mobility simulator study was designed in which 48 participants experienced two different automated mobilities (car and sidewalk mobility). The automation performed based on participants' preferences of how they would like to be driven. A novel aspect of the experiment was that the reliability of the automated mobilities was always maintained. With complete reliability, an effort was made to ensure that the trust could transfer and build across different mobility types. Lack of reliability could lead to decrease of trust, which was avoided to ensure that other dimensions of trust could be influenced which was not the case in our study. It is worth noting that avoiding situations where total reliability is maintained maybe unrealistic.The results showed that participants who drove the car mobility first reported higher dimensions of trust when they later used the sidewalk mobility. In comparison, participants who used the sidewalk mobility first reported lower values across different trust dimensions when they used the car mobility. The findings from the study help inform and identify how people can develop trust in future mobility, and could guide researchers in designing and developing interventions that may help improve the trust and acceptance of future automated mobilities.

\section{Acknowledgements}
The authors wish to thank Anne Mendoza, Elise Ulwelling, Jessie Snyder, and Akhil Modali for their exceptional effort to conduct data collection. Their diligence ensuring compliance with safety protocols helped tremendously with this research study. The authors also wish to thank the employees at Honda Research Institute USA for their invaluable feedback on the study. 

\section{Author contributions}
The authors confirm contribution to the paper as follows: study conception and design: K. Akash, T. Misu, N. Jain, T. Reid, A. Kumar, M. Konishi, J. Hunter;  data collection: K. Akash, Z. Zheng, A. Kumar; analysis and interpretation of results: S. Mehrotra, K. Akash, N. Jain,  M. Konishi, J. Hunter, T. Misu, Z. Zheng;  draft manuscript preparation:  S. Mehrotra, K. Akash, N. Jain;  J. Hunter, T. Reid.

\newpage
\bibliographystyle{trb}
\bibliography{Trust}
\end{document}